\documentclass[reprint,amsmath,amssymb,aps,superscriptaddress,twocolumn,10pt]{revtex4-1}
\usepackage[colorlinks=true,allcolors=blue]{hyperref}
\usepackage{graphicx,color}
\usepackage{dcolumn}
\usepackage{subfigure}
\usepackage{bm}
\usepackage{amsmath,amsfonts,amssymb}
\usepackage{acronym}
\usepackage{enumitem}
\usepackage{array}
\usepackage{multirow}
\allowdisplaybreaks
\newcommand{\be}{\begin{equation}}
\newcommand{\ee}{\end{equation}}
\newcommand{\bea}{\begin{eqnarray}}
\newcommand{\eea}{\end{eqnarray}}

\newcommand{\Msun}{{\rm M}_\odot}

\newcommand{\TRC}{MOE Key Labortory of TianQin Mission,
TianQin Research Center for Gravitational Physics $\&$ School of Physics and Astronomy,
Frontiers Science Center for TianQin,
CNSA Research Center for Gravitational Waves,
Sun Yat-sen University (Zhuhai Campus), Zhuhai 519082, China}
\newacro{EMRI}{Extreme Mass Ratio Inspiral}
\newacro{MBH}{massive black hole}
\newacro{NHT}{no-hair theorem}
\newacro{DWD}{double white dwarf}
\newacro{GW}{gravitational wave}
\newacro{AK}{analytic kludge}
\newacro{NK}{numerical kludge}
\newacro{AAK}{augmented analytic kludge}
\newacro{CO}{compact object}
\newacro{PE}{parameter estimation}
\newacro{SNR}{signal-to-noise ratio}
\newacro{PN}{post newtonion}
\newacro{FIM}{Fisher information matrix}
\newacro{LSO}{last stable orbit}
\newacro{ISCO}{innermost stable circular orbit}
\newacro{BBH}{Binary Black Hole}
\newacro{BNS}{Binary Neutron Star}
\newacro{NS}{Neutron Star}
\begin{document}
\title{Science with the TianQin Observatory: Preliminary Results on Testing the No-hair Theorem with EMRI}
\author{Tieguang Zi}
\affiliation{\TRC}
\author{Jian-dong Zhang}
\email{zhangjd9@mail.sysu.edu.cn}
\affiliation{\TRC}
\author{Hui-Min Fan}
\affiliation{\TRC}
\author{Xue-Ting Zhang}
\affiliation{\TRC}
\author{Yi-Ming Hu}
\email{huyiming@mail.sysu.edu.cn}
\affiliation{\TRC}
\author{Changfu Shi}
\email{shicf6@mail.sysu.edu.cn}
\affiliation{\TRC}
\author{Jianwei Mei}
\email{meijw@sysu.edu.cn}
\affiliation{\TRC}
\date{\today}

\begin{abstract}	
Constituted with a massive black hole and a stellar mass compact object, \ac{EMRI} events hold unique opportunity for the study of massive black holes, such as by measuring and checking the relations among the mass, spin and quadrupole moment of a massive black hole, putting the no-hair theorem to test.
TianQin is a planned space-based gravitational wave observatory and \ac{EMRI} is one of its main types of sources.
It is important to estimate the capacity of TianQin on testing the no-hair theorem with \acp{EMRI}.
In this work, we use the analytic kludge waveform with quadrupole moment corrections and study how the quadrupole moment can be constrained with TianQin.
We find that TianQin can measure the dimensionless quadrupole moment parameter with accuracy to the level of $10^{-5}$ under suitable scenarios.
The choice of the waveform cutoff is found to have significant effect on the result: if the Schwarzschild cutoff is used, the accuracy depends strongly on the mass of the massive black hole, while the spin has negligible impact; if the Kerr cutoff is used, however, the dependence on the spin is more significant.
We have also analyzed the cases when TianQin is observing simultaneously with other detectors such as LISA.
\end{abstract}
\maketitle

\section{Introduction}

Black holes are fascinating objects that may hold the key to quantum gravity and to the grand unification of all interactions in nature. Among the many peculiar results concerning black holes, there is strong theoretical evidence for the hypothesis that classical black holes in general relativity are fully determined by their masses, spins and charges \cite{Israel:1967wq,Israel:1967za,Carter:1971zc,Robinson:1975bv,Hawking:1971tu,Hawking:1971vc} . This hypothesis is known as the \ac{NHT}, and it implicates that classical black holes are astonishingly simple.

Various experimental test of \ac{NHT} has been carried out \cite{Cardoso:2016ryw,Isi:2019aib,Capano:2020dix,Krishnendu:2019tjp,Johannsen:2015mdd,Psaltis:2015uza,Moore:2015bxa,Abdikamalov:2020oci}.
Since the observation of GW150914 \cite{Abbott:2016blz}, about 50 \ac{GW} signals from the merger of stellar mass binary black holes and binary neutron stars have been detected and published \cite{LIGOScientific:2018mvr,Abbott:2020niy}. This has made it more promising to use \ac{GW} observation to test the \ac{NHT}. However, since all the \ac{GW} events detected so far involve source masses of the order 100 $~\Msun$ and less, the capability to test the \ac{NHT} is very limited \cite{Thrane:2017lqn,Isi:2019aib}. To test the \ac{NHT} effectively, we need to detect much heavier \ac{GW} sources, and this requires much larger \ac{GW} detectors that exceed the size of Earth and can only be put in space.

Suitable for this purpose, TianQin is a space-based \ac{GW} observatory planned to launch around 2035 \cite{Luo:2015ght}.
TianQin will be consisted of three drag-free controlled satellites orbiting Earth at radii of about $10^5$ km,
aiming to detect \acp{GW} in the frequency band of $10^{-4}\sim1$ Hz.
The major sources expected for TianQin include inspiral of Galactic compact binaries, inspiral of stellar mass black hole binaries, \acp{EMRI}, merger of \ac{MBH} binaries, and possibly violent processes in the very early universe and exotic sources such as cosmic strings \cite{Kibble:1976sj,Vilenkin:1984ib,Hindmarsh:1994re}. TianQin is expected to provide key information on the astrophysical history of galaxies and black holes, the dynamics of dense star clusters and galactic centers, the nature of gravity and black holes, the expansion history of the universe, and possibly also the fundamental physics related to the early universe
\cite{Wang:2019ryf, Shi:2019hqa,Bao:2019kgt,Huang:2020rjf,Liu:2020eko,Fan:2020zhy}. A summary of the current progress on science and technology of the TianQin project can be found in \cite{Mei:2020lrl}.

When a stellar mass \ac{CO} orbits and finally plunges into a \ac{MBH}, one gets an \ac{EMRI} event.
\ac{EMRI} is one of the most interesting types of sources for a space-based \ac{GW} detector \cite{AmaroSeoane:2007aw,Berry:2019wgg}.
Current study shows that the detection rate of \acp{EMRI} with TianQin ranges from a few to a few hundreds per year,
depending on the astrophysical models used for the estimation \cite{Fan:2020zhy}.
The \ac{CO} can closely orbit the \ac{MBH} for about $10^5$ cycles before the final plunge,
so the \ac{GW} signal contains a plethora of information about the surroundings of the \ac{MBH}.
A black hole in real astrophysical environment loses electric charges fast and can be treated as neutral \cite{Gibbons:1975kk,Goldreich:1969sb,Ruderman:1975ju,Blandford:1977ds}.
So the geometry surrounding the central \ac{MBH} is well approximated by the Kerr metric following the \ac{NHT}.
We will also assume that the environmental effect is negligible and the motion of the \ac{CO} is totally governed by the geometry surrounding the central \ac{MBH}.
With all these assumptions, testing the \ac{NHT} with \acp{EMRI} boils down to test if the \acp{MBH} at the center of \acp{EMRI} are Kerr black holes.

The gravitational field of a localized object can be expanded in terms of multipole moments \cite{Thorne:1980ru,Backdahl:2006ed,Compere:2017wrj}.
The higher multipole moments of a Kerr black hole are fully determined by the mass $M$ and spin $a$ of the black hole \cite{Geroch:1970cd,Hansen:1974zz},
\begin{equation}\label{kerrmoments}
\mathcal{M}_\ell+i\mathcal{S}_\ell=\mathcal{M}(i a)^\ell\,,
\end{equation}
where $\mathcal{M}_\ell$ and $\mathcal{S}_\ell$ are the mass and current multipole moments, respectively.
The odd mass multipole moments and the even current multipole moments vanish due to the equatorial symmetry of the Kerr metric.
By measuring any of the multipole moments with $\ell\geq2$ and comparing with the prediction of (\ref{kerrmoments}), one can check how much the central \ac{MBH} in an \ac{EMRI} may deviate from a Kerr black hole, placing constraints on the \ac{NHT}.

Ryan has pioneered the work of using LISA to extract information on Kerr multipole moments from \ac{EMRI} signals, assuming that the orbits of  stellar mass \acp{CO}  \cite{Ryan:1995wh,Ryan:1997hg} are circular  on the equatorial plane.
By modifying the \ac{AK} \ac{EMRI} waveform with a quadrupole moment correction characterized by a dimensionless quadrupole parameter, Barack et al. \cite{Barack:2003fp,Barack:2006pq} have predicted that LISA can constrain the dimensionless parameter to the level $10^{-4}$ given that the central \ac{MBH} mass takes a certain value.
Babak et al. \cite{Babak:2017tow} have further studied how LISA can constrain the non-Kerr quadrupole moment by using 12 \acp{EMRI} source models.
Using the same source models, Fan et al. \cite{Fan:2020zhy} have assessed the prospect of using TianQin to detect \acp{EMRI} and have also briefly discussed how TianQin can constrain the non-Kerr quadrupole moment.

In this paper, we carry out a more comprehensive study of how well TianQin can test \ac{NHT} with the detection of \acp{EMRI}.
Similar to \cite{Barack:2003fp,Barack:2006pq}, we modify the \ac{AK} waveform with an extra quadrupole moment characterised by a dimensionless parameter $\mathcal{Q}$, then we study how $\mathcal{Q}$ can be constrained with \ac{EMRI} signals detected by TianQin.
We find that the best constraints come from \acp{MBH} with masses at the order $10^{5.5} ~\Msun$, the accuracy of constraint is proportional to the luminosity distance and inverse proportional to the mass of the \ac{CO}, and the parameters such as the eccentricity and angular parameters don't have significant influence on the result. All these are consistent with what is know previously.

Comparing to existing results, our main finding is that the choice of the waveform cutoff has a strong effect on the projected constraints on $\mathcal{Q}$.
We have considered two typical cutoffs used in the literature: the AKS and AKK cutoffs, corresponding to cutting waveforms off at the last stable orbit (LSO) of a Schwarzschild black hole and Kerr black hole, respectively, where CO is captured by the MBH.
When AKS cutoff is used, the level of constraints depends strongly on the mass of the central \ac{MBH}, while the spin has negligible impact;
when AKK cutoff is used, then the dependence on the spin is more distinct.
What's more, constraints from using AKK cutoff is usually orders better than those from using AKS cutoff, due to the obvious reason that more orbital cycles can be accumulated.

The paper is organized as follows.
In Section \ref{method}, we give a brief review about the concept of quadrupole moment and some basic methods about waveform generation and statistics.
Then, we present our results for TianQin and LISA in Section \ref{result}.
Finally, we give a brief summary in Section \ref{summary}.

\section{Method}\label{method}

\subsection{Quadrupole moments of black holes in alternative theories of gravity}\label{quadGravity}

Except for the mass and angular momentum of the central \ac{MBH}, the quadrupole moment is the dominant term among the multipole moments and can leave a distinct imprint on the \ac{EMRI} waveforms.
So the quadrupole moment is the best choice in testing the \ac{NHT}.
Other higher multipole moments of the central \ac{MBH} may also have corrections to the motion of the \ac{CO} and hence the \ac{EMRI} waveforms, but we will not consider them in this paper.

Potential violation of the \ac{NHT} may arise from alternative theories of gravity containing stationary and axisymmetric black hole solutions. These solutions are different from the Kerr metric and the dependence of the quadrupole moments on the masses and spins are different from the prediction of (\ref{kerrmoments}). Some known examples are listed in TABLE~\ref{tab:quadruple}.

\begin{table}[h]\footnotesize
\begin{center}
\begin{tabular}{|l|l|}
\hline
\textbf{Theory} & \textbf{Quadruple Moment} \\
\hline
f(R) theory & $Q_{\rm f(R)}=\sqrt{(M^2-q^2)}a^2$ \cite{Suvorov:2015yfv}\\
Scalar-tensor theory & $Q_{\rm ST}=\frac{1}{3}m\omega_{S}(1+\omega_{S})$ \cite{Pappas:2014gca}\\
EdGB theory & $Q_{\rm EGB}=-M_2+\big[\frac{1}{3}+\frac{4D_1}{3M^2}+\frac{q^2}{12M^2}\big]M^3$ \cite{Kleihaus:2014lba}\\
bumpy BH & $Q_{\rm bumpy}=-a^2M-\frac{1}{2}\sqrt{\frac{5}{\pi}}B_2M^3$ \cite{Vigeland:2010xe}\\
Kerr-NUT BH & $Q_{\rm NUT}=-(M-i N)a^2$ \cite{Mukherjee:2020how}\\
\hline
\end{tabular}
\caption{The quadrupole moment of stationary and axisymmetric black holes in several modified theories of gravity. See the corresponding references for meaning of parameters.}
\label{tab:quadruple}
\end{center}
\end{table}

\subsection{Contribution of quadrupole moment to EMRI waveforms}\label{quadAKwaveform}

High precision waveforms for \acp{EMRI} with different mass ratios can be obtained with the black hole perturbation theory \cite{Poisson:2011nh}.
The effect of the CO's gravitational field on its own orbits, called the self-force effect, has also been included in the perturbation treatment \cite{Barack:2009ux}.
Due to the complexity of the \ac{CO} motion around the central \ac{MBH}, however, it is still technologically challenging to obtain the waveform of an \ac{EMRI} with enough accuracy and efficiency for actual data analysis.
Currently, there exist several different kinds of waveform models, such as the kludge waveforms which includes \ac{AK} \cite{Barack:2003fp}, \ac{NK} \cite{Gair:2005ih}, and\ac{AAK} \cite{Chua:2016jnd,Chua:2017ujo}, and the recently developed FastEMRIWaveforms \cite{Chua:2020stf,Katz:2021yft}.

In this work, we choose to use the \ac{AK} method, which is also used in the previous analysis for LISA \cite{Barack:2006pq,Babak:2017tow} and TianQin \cite{Fan:2020zhy}.
The \ac{AK} model could describe the main feature of an \ac{EMRI} waveform, and it is much more straightforward to add the corrections of the quadrupole moment $\mathcal{Q}$.
So we perform current study with AK waveforms including quadrupole moment corrections.
However, we should notice that due to the mismap of the orbital frequencies, the frequencies in the AK model is overall too high, as pointed out in \cite{Chua:2017ujo}.
Thus it will result in a non-negligible bias in the matched filtering if we use AK waveform.
But we also need to mention that a quadrupole moment included waveform based on the \ac{AAK} method is recently developed in \cite{Liu:2020ghq}.
The results of that work indicate that the \ac{PE} accuracy will not
be seriously influenced by the choice of waveform models.
The \ac{PE} accuracy achieved by the so called QAK and QAAK therein
are almost the same by orders of magnitude.
Thus it's still reasonable to use the \ac{AK} method in this work.

In the \ac{AK} method, the \ac{EMRI} waveform  is described by 14 parameters, not considering the spinning \ac{CO},
\begin{align}
\lambda^i \equiv~ &(\lambda^1,\cdots, \lambda^{14}) \notag\\
 \quad =\Big[&t_0,\ln\mu, \ln M,\hat{S},e_{\rm LSO}, \tilde{\gamma}_0,\Phi_0,\cos\theta_S, \nonumber\\
 &\phi_S,\cos\lambda,\alpha_0, \cos\theta_K,\phi_K,\ln(\mu/D)\Big]\,,
\end{align}
where definition of the parameters can be found in \cite{Barack:2003fp}.
Violation of the \ac{NHT} can be introduced through the dimensionless quadrupole parameter $\mathcal{Q}$:
\begin{equation}\label{quadrupole}
\mathcal{Q} \equiv Q/M^3\,,
\end{equation}
where $Q=-Ma^2$ and $\mathcal{Q}=-a^2/M$ correspond to no violation of the \ac{NHT}.
Following \cite{Barack:2003fp,Barack:2006pq}, we use Post-Newtonian equations, see appendix \ref{evolution}, of the orbital phase angles $(\Phi,\tilde{\gamma},\alpha)$ and of the frequency and eccentricity $(\nu,e)$ to determine the dynamics of the \ac{CO}.

At the final stage of \ac{EMRI}, when the \ac{CO} passed the boundary of stable orbits,
it will plunge into the \ac{MBH} directly in a short time.
So, we need to introduce a cutoff to the \ac{AK} waveform.
For a \ac{CO} moving in the equator plane of the central \ac{MBH}, the cutoff is usually taken to be the \ac{LSO}.
The orbital frequency reaches to the maximum value at the \ac{LSO} \cite{Cutler:1994pb},
\begin{equation}
\nu_\text{LSO} = \frac{1}{2\pi M}\Big( \frac{1-e^2}{r_\text{ISCO}/M+2e}\Big)^{3/2},
\end{equation}
where $r_\text{ISCO}$ is the radius of \ac{ISCO}.

When the central \ac{MBH} is a Schwarzschild black hole, $r_\text{ISCO}=6M$ \cite{Cutler:1994pb}, and the cutoff is shortened as the AKS cutoff.
When the central \ac{MBH} is a Kerr black hole, we have the AKK cutoff, obtained for prograde orbits \cite{Bardeen:1972fi}:
\begin{align}\label{akk}
&r_\text{ISCO}/M=3+z_2-\sqrt{[(3-z_1)(3+z_1+2z_2)]}\,, \nonumber\\
&z_1=1+(1-\hat{S}^2)^{1/3}[(1+\hat{S})^{1/3}+(1-\hat{S})^{1/3}]\,,\nonumber\\
&z_2= \sqrt{(3\hat{S}^2+z_1^2)}\,.
\end{align}
And for retrograde orbit, the plunge happens very far from the \ac{MBH},
and thus the gravitational radiation will be too weak to be detected.
According to the result of \cite{Fan:2020zhy}, most of the detected events have prograde orbits.
The AKK cutoff is closer to the \ac{MBH} than the AKS cutoff.
So the AKK cutoff always generate more optimistic result than the AKS cutoff.
However, this dosen't mean the plunge has to happen between this two criteria,
since this is only true for equatorial orbit.

A recent work \cite{Stein:2019buj} aims to find out a more realistic value for the \ac{LSO}, but the result obtained for these two criteria can still show some fundamental features.
We will see in the next section that the two cutoffs lead to drastically different results,
both in terms of the level of constraints predicted and of the dependence of the constraints on the source parameters.

\subsection{Constraints on the quadrupole moment}\label{FIMways}

For a signal with a large \ac{SNR}, the statistical uncertainties in the parameters are approximated by
\begin{equation}
\Delta\lambda^i \approx \Big[(\Gamma^{-1})^{ii}\Big]^{1/2}\,,
\end{equation}
where $\Gamma$ is the \ac{FIM} whose elements are defined through
\begin{equation}
\Gamma_{ij}\equiv \Big(\frac{\partial h}{\partial\lambda^i} \Big|\frac{\partial h}{\partial\lambda^j}\Big)\,.
\end{equation}
Here the inner product is defined as \cite{Cutler:1994ys}
\begin{equation}
(g|h) = 2 \int^{f_{high}}_{f_{low}} \frac{g^*(f)h(f)+g(f)h^*(f)}{S_n(f)}df\,,
\end{equation}
where $f_{low,high}$ denote the detector-dependent lower and upper truncation frequencies and $S_n(f)$ is the sensitivity of the detector.
For TianQin \cite{Luo:2015ght}, we have:
\begin{align}
S_n(f)~=~ &\frac{1}{L^2_0}\Big[\frac{4S_a}{(2\pi f)^4}\Big( 1+\frac{10^{-4} {\rm Hz} }{f} \Big)+ S_x\Big]\nonumber\\
&\times\Big[1+\Big(\frac{2f L_0}{0.41c}\Big)^2\Big],
\end{align}
where $L_0=\sqrt{3}\times 10^5$ km is the arm length ,
$S_a=1\times10^{-30}{\rm ms^{-4}Hz^{-1}}$ and \(S_x=1\times10^{-24}~{\rm m^2Hz^{-1}}\) are the power densities of the residual acceleration
on each test mass and the displacement measurement noise in a single laser link, respectively.

\section{Results}\label{result}

Taking the central value of $\mathcal{Q}$ to be that of Kerr \cite{Babak:2017tow},
we estimate the constraints that can be imposed on the possible deviations $\Delta\mathcal{Q}$.

We find that $\Delta\mathcal{Q}$ has nearly linear dependence on $t_0$, $D$ and $\mu$, while the result barely depends on the eccentricity $e$ or the angular parameters.
So these parameters are held fixed in our calculations:
$t_0=5$ years,
$D=2$ Gpc,
$\mu=18~{\rm M}_\odot$,
$e=0.1$,
$\lambda=\pi/3$,
$\tilde{\gamma}_0=5\pi/6$,
$\alpha_0=4\pi/5$,
$\theta_S=\pi/5$,
$\phi_S=\pi/4$,
$\theta_K=2\pi/3$,
$\phi_K=3\pi/4$, and
$\Phi_0=\pi/3$.
For $M$, the mass of the \ac{MBH}, we take values from the range $10^5-10^7~\Msun$, based on the result of \cite{Fan:2020zhy}. For $\hat{S}$, the dimensionless spin of the \ac{MBH}, we take values from the range $0\sim0.98$.

\begin{figure*}
	\centering
	\includegraphics[width=0.45\textwidth]{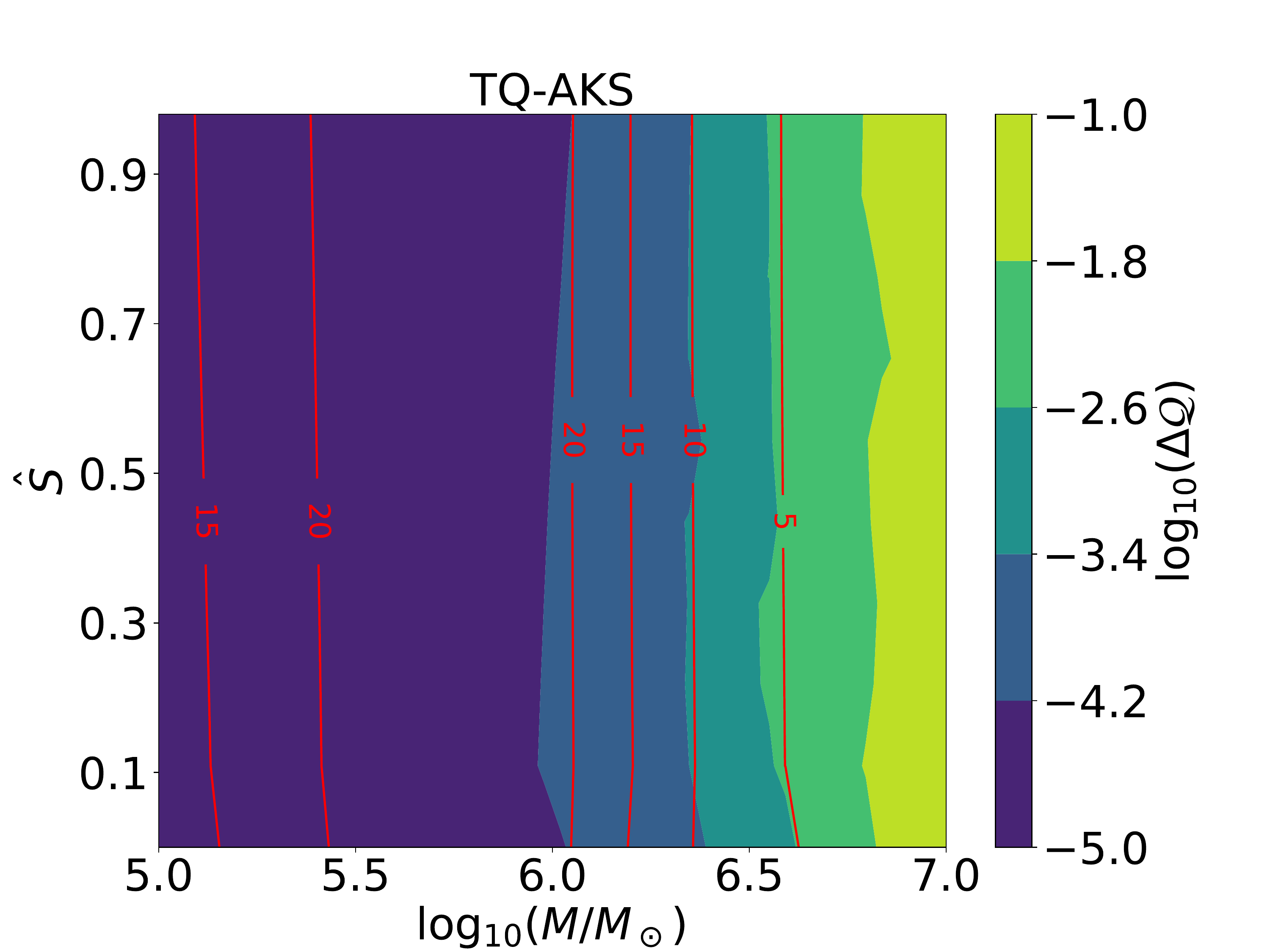}
	\includegraphics[width=0.45\textwidth]{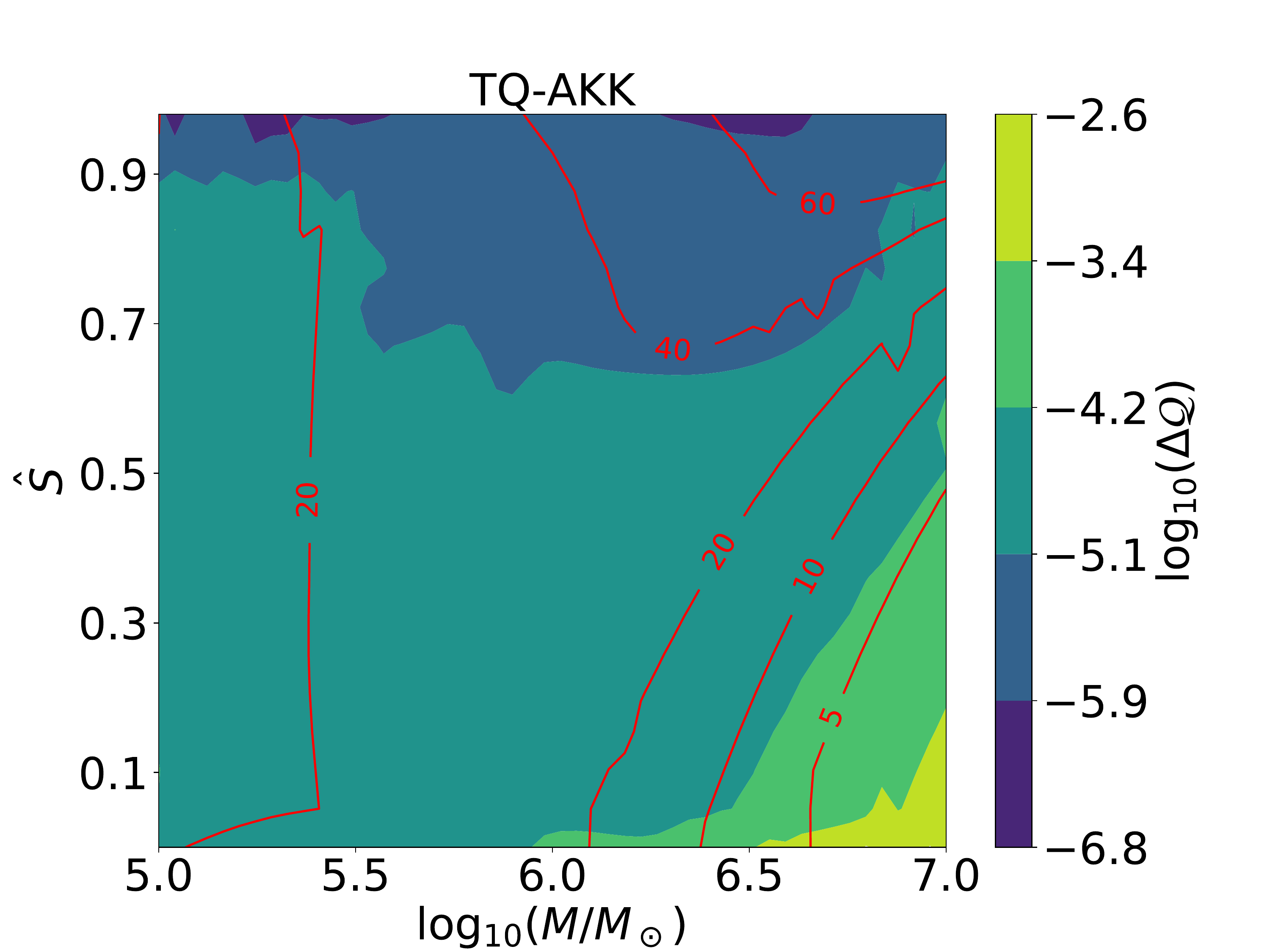}
	\caption{Dependence of $\Delta\mathcal{Q}$ on $M$ and $\hat{S}$ by using \acp{EMRI} detected by TianQin. The left (right) figure corresponds to using AKS (AKK) cutoff. The red curves corresponding to the contour of SNR.}
	\label{TQcontfig1}
\end{figure*}

\begin{table*}[!htbp]
	\caption{$\Delta\mathcal{Q}$ for TQ with different mass $M$ and spin $\hat{S}$ of the \ac{MBH},
		the plain and bold values correspond to AKS and AKK cutoff respectively.}\label{TQtable}
	\begin{center}
		\setlength{\tabcolsep}{7.5mm}
		\begin{tabular}{|c|c|c|c|c|c|}
			\hline
			\multirow{2}{*}{$\hat{S}$}& \multicolumn{5}{c|}
			{MBH mass $\log_{10}(M/~\Msun)$}\\
			\cline{2-6}
			& $5.0$ &$5.5$ &$6.0$ &$6.5$ & $7.0$ \\\hline
			$0.1$ & $6.6\times10^{-5}~~ $ &$2.4\times10^{-5}$ &$8.3\times10^{-5}$
			&$1.6\times10^{-3}$ &$7.5\times10^{-2}$\\
			&$\boldsymbol{6.4\times10^{-5}}$ &$\boldsymbol{2.3\times10^{-5}}$
			&$\boldsymbol{4.1\times10^{-5}}$ &$\boldsymbol{6.8\times10^{-5}}$
			&$\boldsymbol{6.3\times10^{-4}}$\\
			\hline
			$0.2$ &$6.4\times10^{-5}$ &$2.5\times10^{-5}$& $7.8\times10^{-5}$
			&$1.5\times10^{-3}$&$7.8\times10^{-2}$\\
			&$\boldsymbol{5.3\times10^{-5}}$ &$\boldsymbol{2.1\times10^{-5}}$
			&$\boldsymbol{3.4\times10^{-5}}$ &$\boldsymbol{6.3\times10^{-5}}$
			&$\boldsymbol{3.8\times10^{-4}}$ \\
			\hline
			$0.3$ &$6.6\times10^{-5}$ &$2.2\times10^{-5} $&$7.3\times10^{-5} $
			&$1.7\times10^{-3}$ & $7.6\times10^{-2}$\\
			&$\boldsymbol{3.2\times10^{-5}}$ &$\boldsymbol{1.6\times10^{-5}}$
			&$\boldsymbol{2.8\times10^{-5}}$ &$\boldsymbol{4.1\times10^{-5}}$
			&$\boldsymbol{2.2\times10^{-4}}$ \\
			\hline
			$0.4 $&$ 6.6\times10^{-5} $&$2.2\times10^{-5} $&$7.3\times10^{-5}
			$&$1.8\times10^{-3}$&$ 6.8\times10^{-2}$\\
			
			&$\boldsymbol{2.1\times10^{-5}}$ &$\boldsymbol{1.5\times10^{-5}}$
			&$\boldsymbol{2.1\times10^{-5}}$ &$\boldsymbol{3.2\times10^{-5}}$
			&$\boldsymbol{1.2\times10^{-4}}$ \\
			\hline
			$0.5$&$5.8\times10^{-5} $&$2.3\times10^{-5} $&$7.1\times10^{-5}$
			&$2.2\times10^{-3} $&$8.3\times10^{-2}$\\
			&$\boldsymbol{1.8\times10^{-5}}$ &$\boldsymbol{1.2\times10^{-5}}$
			&$\boldsymbol{1.7\times10^{-5}}$ &$\boldsymbol{2.1\times10^{-5}}$
			&$\boldsymbol{1.1\times10^{-4}}$ \\
			\hline
			$0.6 $&$ 6.9\times10^{-5} $&$2.3\times10^{-5} $&$7.1\times10^{-5}
			$&$2.2\times10^{-3}$ &$8.3\times10^{-2}${\tiny }\\
			&$\boldsymbol{1.6\times10^{-5}}$ &$\boldsymbol{9.5\times10^{-6}}$
			&$\boldsymbol{1.1\times10^{-5}}$ &$\boldsymbol{1.4\times10^{-5}}$
			&$\boldsymbol{6.4\times10^{-5}}$ \\
			\hline
			$0.7$&$6.5\times10^{-5} $&$2.4\times10^{-6} $&$6.5\times10^{-5}$
			&$9.5\times10^{-4}$&$4.6\times10^{-2}$\\
			&$\boldsymbol{1.5\times10^{-5}}$ &$\boldsymbol{6.2\times10^{-6}}$
			&$\boldsymbol{7.5\times10^{-6}}$ &$\boldsymbol{7.8\times10^{-6}}$
			&$\boldsymbol{5.8\times10^{-5}}$ \\
			\hline
			$0.8$&$ 5.5\times10^{-5} $&$2.3\times10^{-5} $&$6.2\times10^{-5}$
			&$2.3\times10^{-3}$ &$7.7\times10^{-2}$\\
			&$\boldsymbol{1.3\times10^{-5}}$ &$\boldsymbol{4.0\times10^{-6}}$
			&$\boldsymbol{5.5\times10^{-6}}$ &$\boldsymbol{5.1\times10^{-6}}$
			&$\boldsymbol{5.2\times10^{-5}}$ \\
			\hline
			$0.9$&$ 5.2\times10^{-5} $&$2.3\times10^{-5} $&$5.9\times10^{-5} $
			&$2.2\times10^{-3}$&$8.1\times10^{-2}$\\
			&$\boldsymbol{1.2\times10^{-5}}$ &$\boldsymbol{2.2\times10^{-6}}$
			&$\boldsymbol{4.8\times10^{-6}}$ &$\boldsymbol{4.3\times10^{-6}}$
			&$\boldsymbol{4.1\times10^{-5}}$ \\
			\hline
		\end{tabular}
	\end{center}
\end{table*}

\begin{figure}[!htbp]
	\centering
	\includegraphics[width=0.48\textwidth]{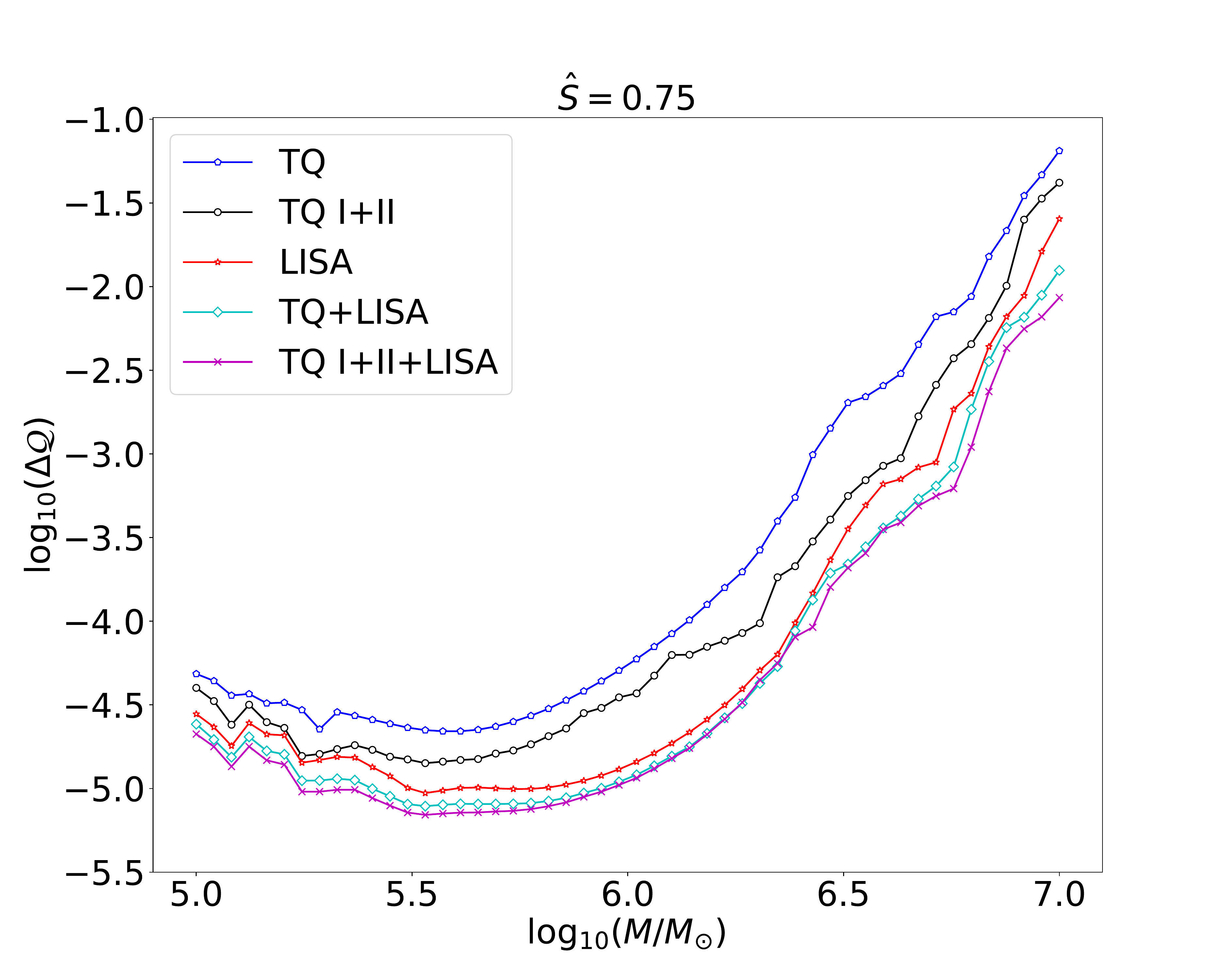}
	\caption{Dependence of $\Delta\mathcal{Q}$ on $M$, with $\hat{S}=0.75$ using the AKS waveform with different detector configurations. The remaining parameters are the same as in FIG. \ref{TQcontfig1}.}
	\label{Q-MBH-fixspin-AKS}
\end{figure}

The constraints obtained with both AKS and AKK cutoffs, using \acp{EMRI} signals that can be detected with TianQin, are illustrated in FIG. \ref{TQcontfig1} and listed in TABLE \ref {TQtable}.
We also plotted the contour of \ac{SNR} with red curve,
and one can find that the result is strongly correlated with the value of SNR.
One can also see that there is drastic difference between the results obtained with the two cutoffs:
\begin{itemize}
\item In the chosen range for the mass $M$ and the spin parameter $\hat{S}$, constraints achievable with the AKS cutoff is in the range $10^{-1}\sim10^{-5}$ and those from the AKK cutoff is in the range $10^{-2.6}\sim10^{-6.8}$,
    with the latter being one to two orders better than the former.
\item With the AKS cutoff, the constraints is dominated by the mass of \ac{MBH}, while the effect of spin can be neglected. The dependence of $\Delta \mathcal{Q}$ on mass is plotted in FIG. \ref{Q-MBH-fixspin-AKS}. The best constraint, at the level $\Delta\mathcal{Q}\sim 10^{-5}$ is achievable with $M\sim 10^{5.5}~\Msun$.
\item With the AKK cutoff, the constraints depend most significantly on the spin parameter of the \ac{MBH}, showing a general trend that larger spin leads to more stringent constraints on $\mathcal{Q}$.
\end{itemize}
The difference is likely due to the fact that the Kerr \ac{LSO} is closer to the \ac{MBH} than the Schwarzschild \ac{LSO},
especially when there is a large spin.

\begin{figure}[htbp]
	\centering
	\includegraphics[width=0.48\textwidth]{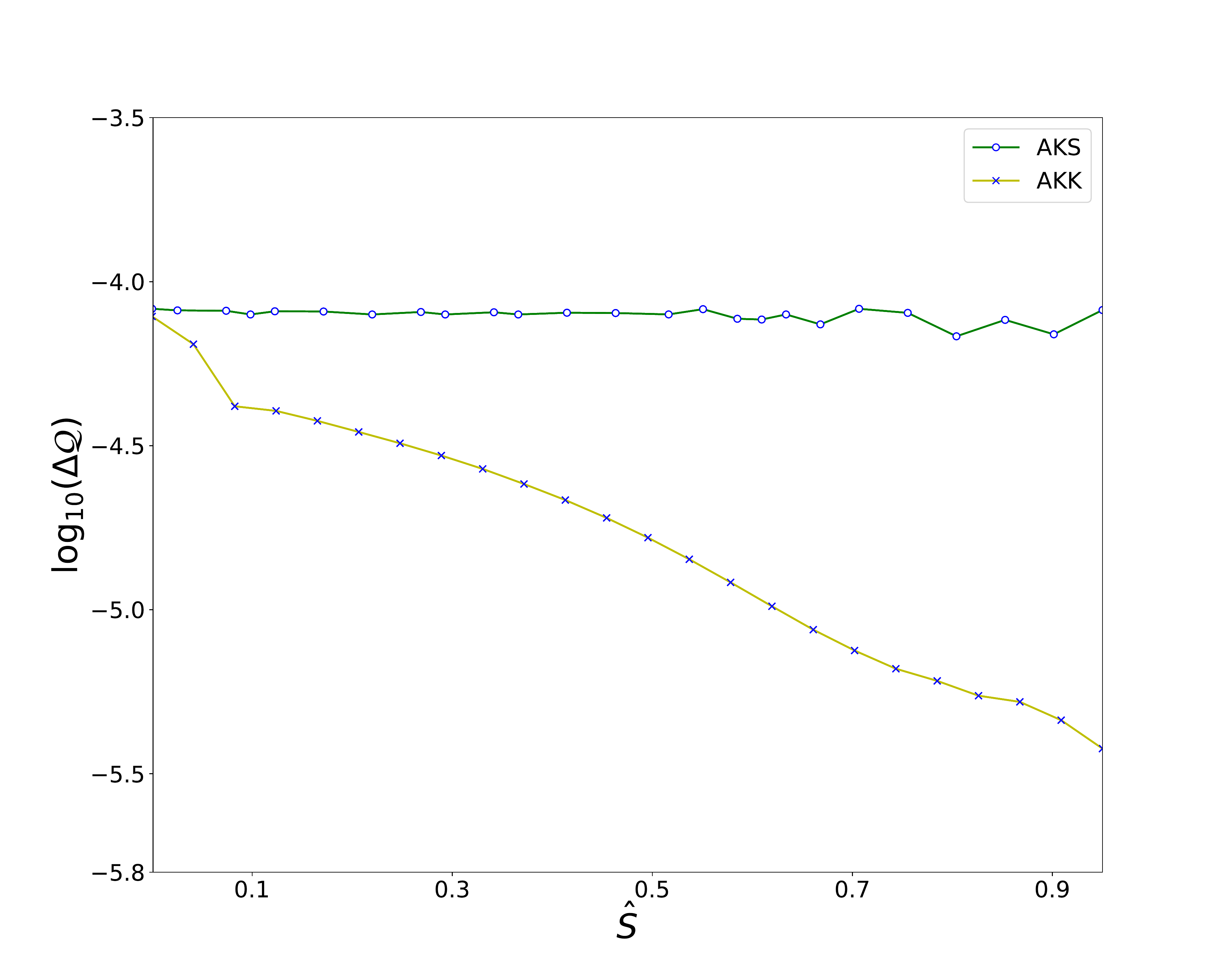}
	\caption{Dependence of $\Delta\mathcal{Q}$ on the spin parameter $\hat{S}$, assuming $M=10^6~\Msun$.}\label{TQIquad-spin3}
\end{figure}

The dependence of $\Delta\mathcal{Q}$ on the spin parameter $\hat{S}$ is illustrated in FIG. \ref{TQIquad-spin3} for the \ac{MBH} mass of $10^6~\Msun$.
We see that $\Delta\mathcal{Q}$ changes very little with varying $\hat{S}$ in the case with the AKS cutoff,
but decreases steadily with growing $\hat{S}$ in the case with the AKK cutoff.

\begin{figure}
	\centering
	\includegraphics[width=0.48\textwidth]{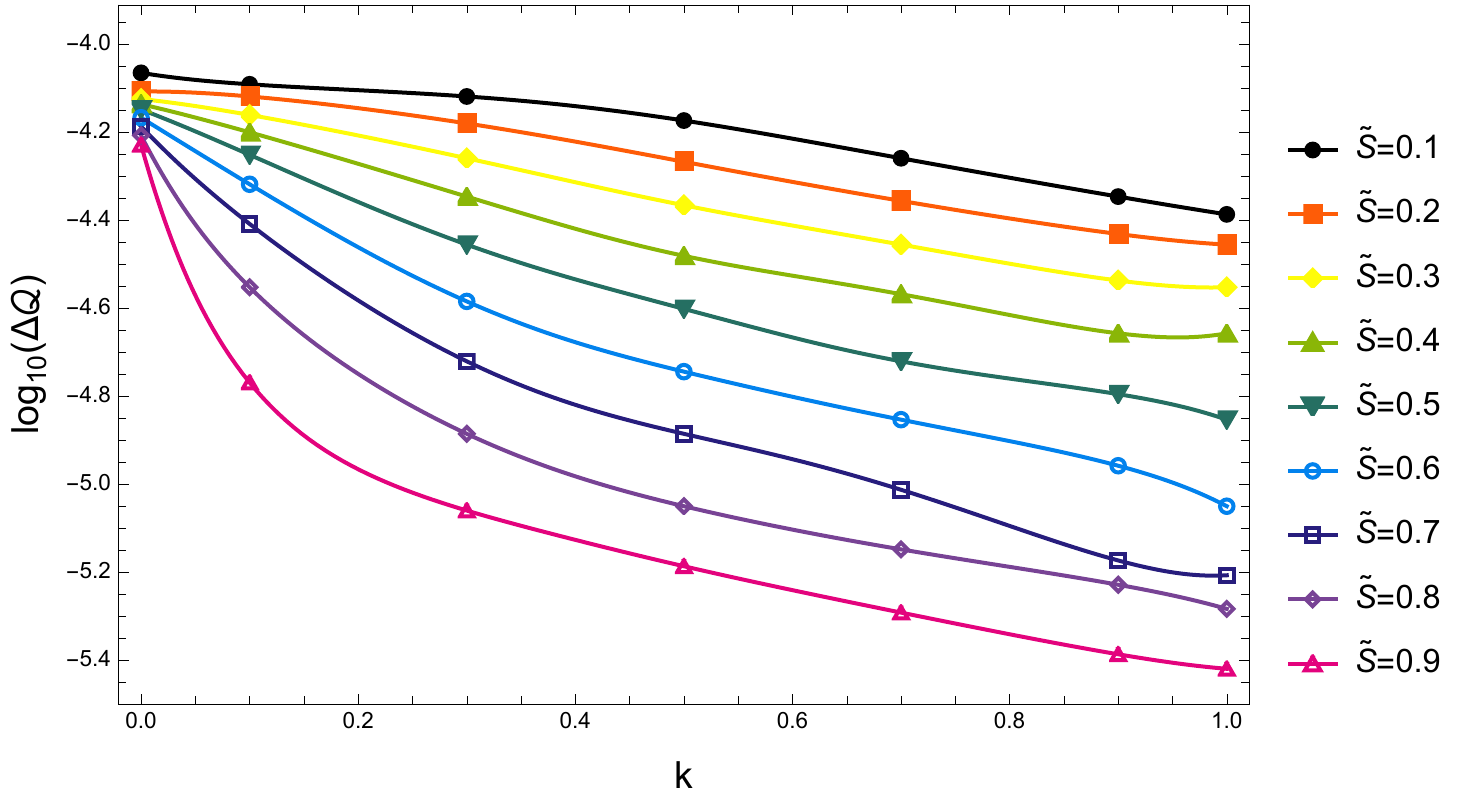}
	\caption{Dependence of $\Delta\mathcal{Q}$ on $\hat{S}$ and $k$, assuming $M=10^6~\Msun$.}
	\label{nuocutoff}
\end{figure}

To better understand the dependence on the cutoff, we introduce a new cutoff interpolating AKS cutoff and AKK cutoff,
\begin{equation}
\nu_k=\nu_{\rm AKK}+k(\nu_{\rm AKK}-\nu_{\rm AKS})\,,\quad k\in[0,1]\,.
\end{equation}
The dependence of $\Delta\mathcal{Q}$ on $\hat{S}$ and $k$ is illustrated in FIG. \ref{nuocutoff} with $M=10^6~\Msun$.
One can see that the dependence on $\hat{S}$ becomes more and more significant as $k$ varies from 0 to 1.
This treatment may not have practical significance, since it has nothing to do with the realistic plunge.
But it indicates the fact that a better understanding of when plunge happens is needed for more precise study,
and an extrapolation can tell us the tendency of what will happen if the plunge happens outside these crude boundries.

\begin{figure}
	\centering
	\includegraphics[width=0.48\textwidth]{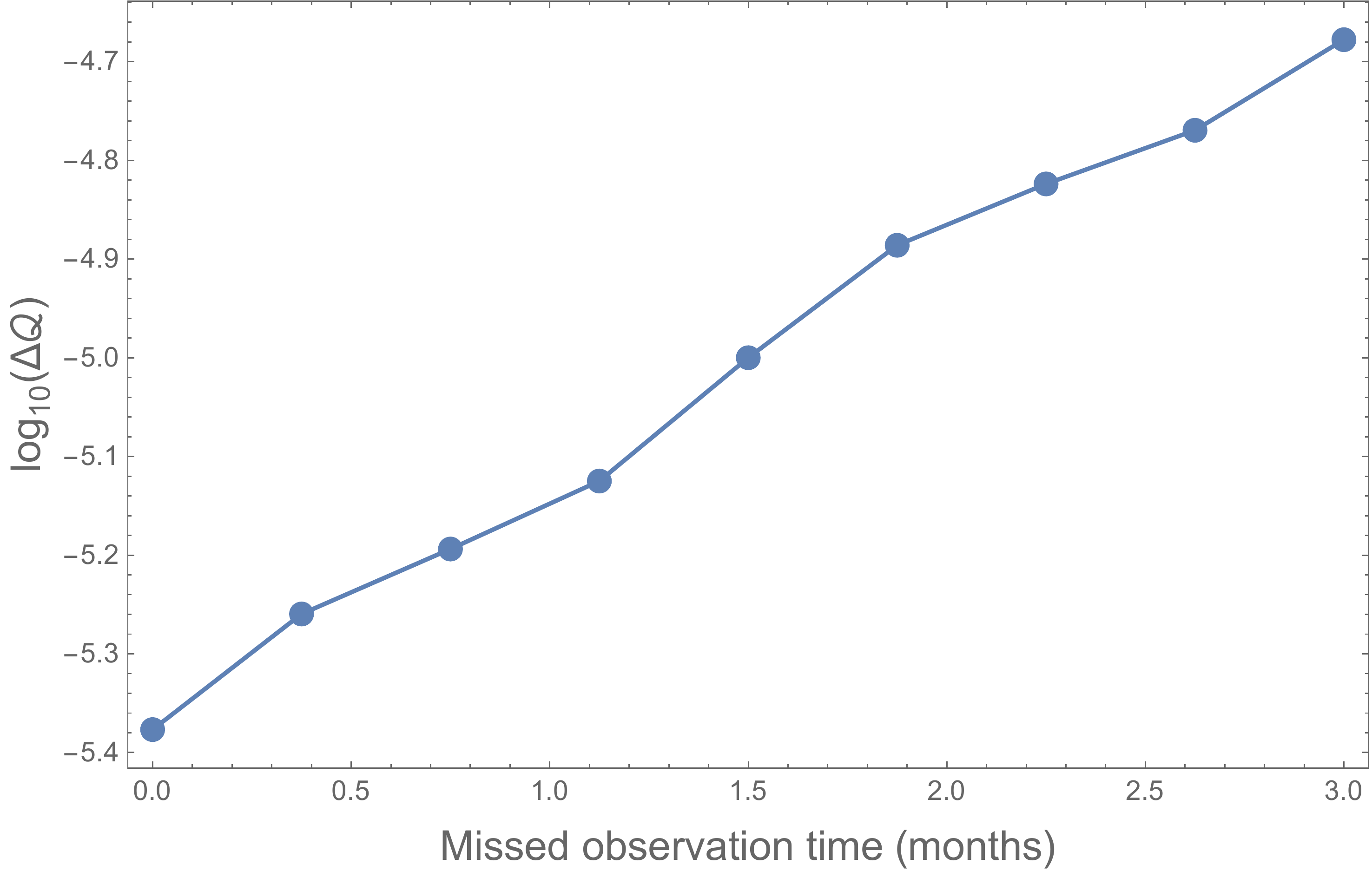}
	\caption{Dependence of $\Delta\mathcal{Q}$ on the missing observation time before plunge, assuming $M=10^6~\Msun$ and $\hat{S}=0.8$.}
	\label{3month}
\end{figure}

The baseline concept of TianQin adopts a ``3 month on + 3 month off'' observation scheme. If the plunge of the \ac{CO} happens at the gap between two observation windows, some data at the final stage of the \ac{EMRI} will be lost.
To assess the impact of this loss of data on science, we plot in FIG. \ref{3month} the dependence of $\Delta\mathcal{Q}$ on the amount of time with lost observation for the final stage of an \ac{EMRI} event.
In the worst case scenario, when totally three month of data are lost for the final stage of an \ac{EMRI}, the constraint can be worsen by as large as about 5 times.

For completeness, we have also studied the constraints on $\mathcal{Q}$ with a variety of detector and detector networks,
such as LISA (FIG. \ref{LISAcontfig4}), TQ I+II (FIG. \ref{TQI+IIcontfig2}),
TQ + LISA (FIG. \ref{TQ+LISAcontfig5}), and TQ I+II + LISA (FIG. \ref{TQI+II+LISAcontfig2}).
A detailed explanation of the aforementioned detector networks can be found in \cite{Huang:2020rjf}.

To illustrate the result more clearly, we also plot the dependence of $\Delta\mathcal{Q}$ on the mass $M$ for fixed spin in the case with AKS cutoff.
As we described above, the spin parameter will not influence the result significantly, so we just plot for $\hat{S}=0.75$ in FIG. \ref{Q-MBH-fixspin-AKS}.

\section{SUMMARY AND FUTURE WORK}\label{summary}

In this paper we have presented a preliminary study of testing \ac{NHT} with \ac{EMRI} using the TianQin observatory.

With the dimensionless parameter $\mathcal{Q}$ to characterize the \ac{NHT} violation,
we have put the bound on such parameter using the AK waveform with quadrupole corrections.
One can constrain $\mathcal{Q}$ to about $10^{-5}$ level with a 5 year \ac{EMRI} observation by TianQin.
We also calculate the situation about joint detection with LISA.
The results show that the \ac{PE} of quadruple moment can be improved several times or more by joint detection compared with individual observatory.
Since these results are obtained by the inaccurate \ac{AK} waveform,
the result would be updated  in the future if some more accurate waveforms with quadrupole moment correction are ready to be used.
However, we do not expect the results would significantly change.

We also find that the choice of plunge will influence the results significantly, especially for higher spin \ac{MBH}.
While the \ac{MBH}s usually trend to fast spin for its astrophysical grown process,
the choices of the roughly Kerr or Schwarzchild cutoff are not accuracy enough to obtain exact and reliable results.
It seems quite important to analyze the physical meaningful plunge cutoff in detail,
and we will try to include the more realistic consideration of plunge in the future work.

In addition, EMRI waveform is strongly influenced by multipole moment of MBH to some degree,
for the purpose of obtaining more plentiful information on the multipolar structure of MBH and testing NHT,
one need more reliable \ac{EMRI} waveform model.

\begin{acknowledgments}
The authors thank L.Barack  for the helpful communication. This work has been supported by the Guangdong Major Project of Basic and Applied Basic Research (Grant No. 2019B030302001), the National Key Research and Development Program of China (No. 2020YFC2201400), the Natural Science Foundation of China (Grants No. 11805286, No. 11690022), the China Postdoctoral Science Foundation  (Grant No. 2020M683016), and the Guangdong Basic and Applied Basic Research Foundation(Grant No. 2021A1515010319). This work is supported by National Supercomputer Center in Guangzhou.
\end{acknowledgments}

\appendix
\section{Construction of Waveform}

In the \ac{AK} method, the equation for orbital evolution is:
\begin{align}\label{evolution}
\dot{\Phi}~=~&2\pi \nu, \nonumber\\
\dot{\nu}~=~&\frac{48\mu}{5\pi M^3}(2\pi M\nu)^{11/3}(1-e^2)^{-9/2} \Big\{\Big[1+\frac{73}{24}e^2+\frac{37}{96}e^4\Big] \nonumber\\
&\times(1-e^2)+(2\pi M\nu)^{2/3}\Big[\frac{1273}{336}-\frac{2561}{336}e^2-\frac{3885}{128}e^4 \nonumber\\
&-\frac{13147}{5376}e^6\Big] - \hat{S} \cos\lambda (2\pi M\nu)(1-e^2)^{-1/2}\Big[\frac{73}{12}\nonumber\\
&+\frac{1211}{24}e^2+\frac{3143}{96}e^4+\frac{65}{64}e^6\Big] -\mathcal{Q}(2\pi M\nu)^{4/3}\nonumber\\
&\times(1-e^2)^{-1}\Big[\frac{33}{16}+\frac{359}{32}e^2-\frac{527}{96}\sin^2\lambda\Big]\Big\}, \nonumber\\
\dot{e}~=~&-\frac{e\mu}{15M^2}(2\pi M\nu)^{8/3}(1-e^2)^{-7/2}\big[(304+121e^2) \nonumber\\
&\times(1-e^2)(1+12(2\pi M\nu)^{2/3}) -\frac{1}{56}(2\pi M\nu)^{2/3} \nonumber\\
&\times(133640+108984e^2+25211e^4)\big]
\nonumber\\&+\frac{e \mu}{M^2} \hat{S}\cos\lambda (2\pi M\nu)^{11/3}(1-e^2)^{-4}
\nonumber\\& \times\Big[\frac{1364}{5}+\frac{5032}{15}e^2
+\frac{263}{10}e^4)\Big],\nonumber\\
\dot{\tilde{\gamma}}~ =~& \frac{3}{2}\pi \nu(2\pi M\nu)^{2/3}(1-e^2)^{-1}\Big[4+(2\pi M\nu)^{2/3}(1-e^2)^{-1} \nonumber\\
&\times(26-15e^2)\Big] - 12\pi\nu\hat{S}\cos\lambda (2\pi M\nu)(1-e^2)^{-3/2} \nonumber\\
& - \frac{3}{2}\pi\nu \mathcal{Q} (2\pi M\nu)^{4/3}(1-e^2)^{-2}(5\cos\lambda-1), \nonumber\\
\dot{\alpha}~=~&4\pi\nu\hat{S}(2\pi M\nu)(1-e^2)^{-3/2}+3\pi\nu\mathcal{Q}\cos\lambda\nonumber\\
&\times(2\pi M\nu)^{4/3}(1-e^2)^{-2},\nonumber\\
\end{align}
where dot denotes the derivative with respect to time.

The waveform of the two polarizations are defined via an n-harmonic waveform:
\begin{align}\label{amplitude}
h_+ \equiv \sum_n A_n^+ = \sum_n &-\Big[1+(\hat{L}\cdot\hat{n})^2\Big]\Big[a_n \cos2\gamma -b_n\sin2\gamma\Big]\nonumber\\
& +c_n\Big[1-(\hat{L}\cdot\hat{n})^2\Big], \nonumber\\
h_\times \equiv \sum_n A_n^\times = \sum_n &2(\hat{L}\cdot\hat{n})\Big[b_n\cos2\gamma+a_n\sin2\gamma\Big].
\end{align}
It's determined by the position of the source $\hat{n}$, and the direction of the orbital angular momentum $\hat{L}$. The coefficients $(a_n, b_n, c_n)$ is determined by the eccentricity $e$ and mean anomaly $\Phi$, as given by Peter and Mathews \cite{Peters:1963ux}
\begin{align}
a_n =~ &-n \mathcal{A} \Big[J_{n-2}(ne)-2eJ_{n-1}(ne)+\frac{2}{n}J_n(ne) \nonumber\\
& +2J_{n+1}(ne) -J_{n+2}(ne)\Big]\cos(n\Phi), \nonumber\\
b_n =~ &-n \mathcal{A}(1-e^2)^{1/2}\Big[J_{n-2}(ne)-2J_{n}(ne)+J_{n+2}(ne)\Big] \nonumber\\
 &\times\sin(n\Phi),\nonumber\\
c_n =~& 2\mathcal{A}J_n(ne)\cos(n\Phi),\nonumber\\
\mathcal{A} = ~& (2\pi M\nu )^{2/3}\mu/D,
\end{align}
where the $J_n$ is Bessel functions of the first kind.

Since the equilateral triangle detectors such as TianQin can be used to construct two independent Michelson interferometers,
the signal responded by such two interferometers can be written as:
\begin{eqnarray}
h_{I,II} = \frac{\sqrt{3}}{2} \Big(F^+_{I,II}h^+ + F^\times_{I,II}h^\times \Big)
\end{eqnarray}
where the antenna pattern functions $F^{+,\times}_{I,II}$ \cite{Cutler:1994ys} of detector depend on the orbits of satellites. Detail information of TianQin respond function for \ac{EMRI} signal can be found in \cite{Fan:2020zhy}.

\bibliographystyle{apsrev4-1}
\bibliography{TQ-EMRI-TEST}

\begin{figure*}[!htbp]
\centering
\includegraphics[width=0.45\textwidth]{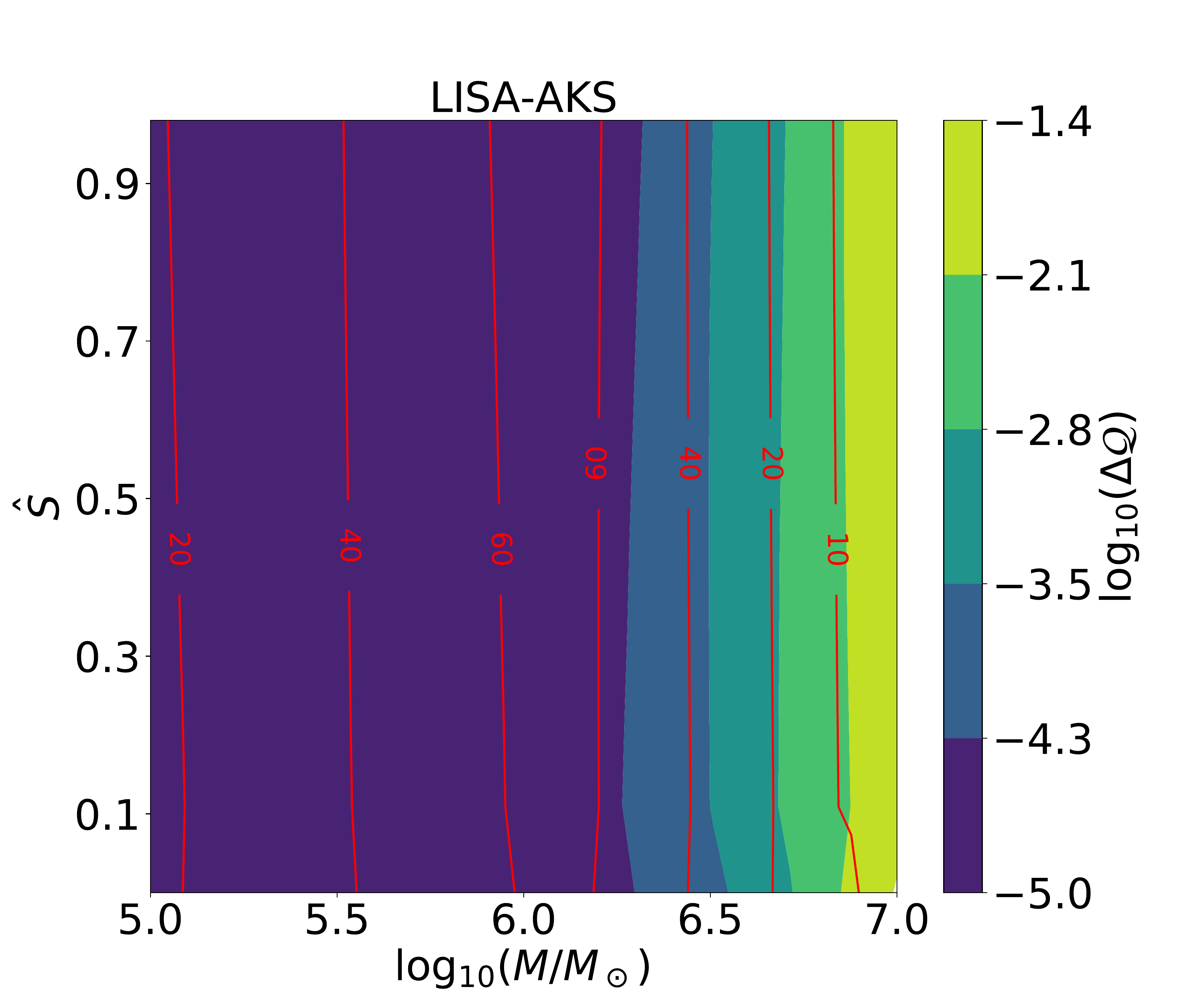}
\includegraphics[width=0.45\textwidth]{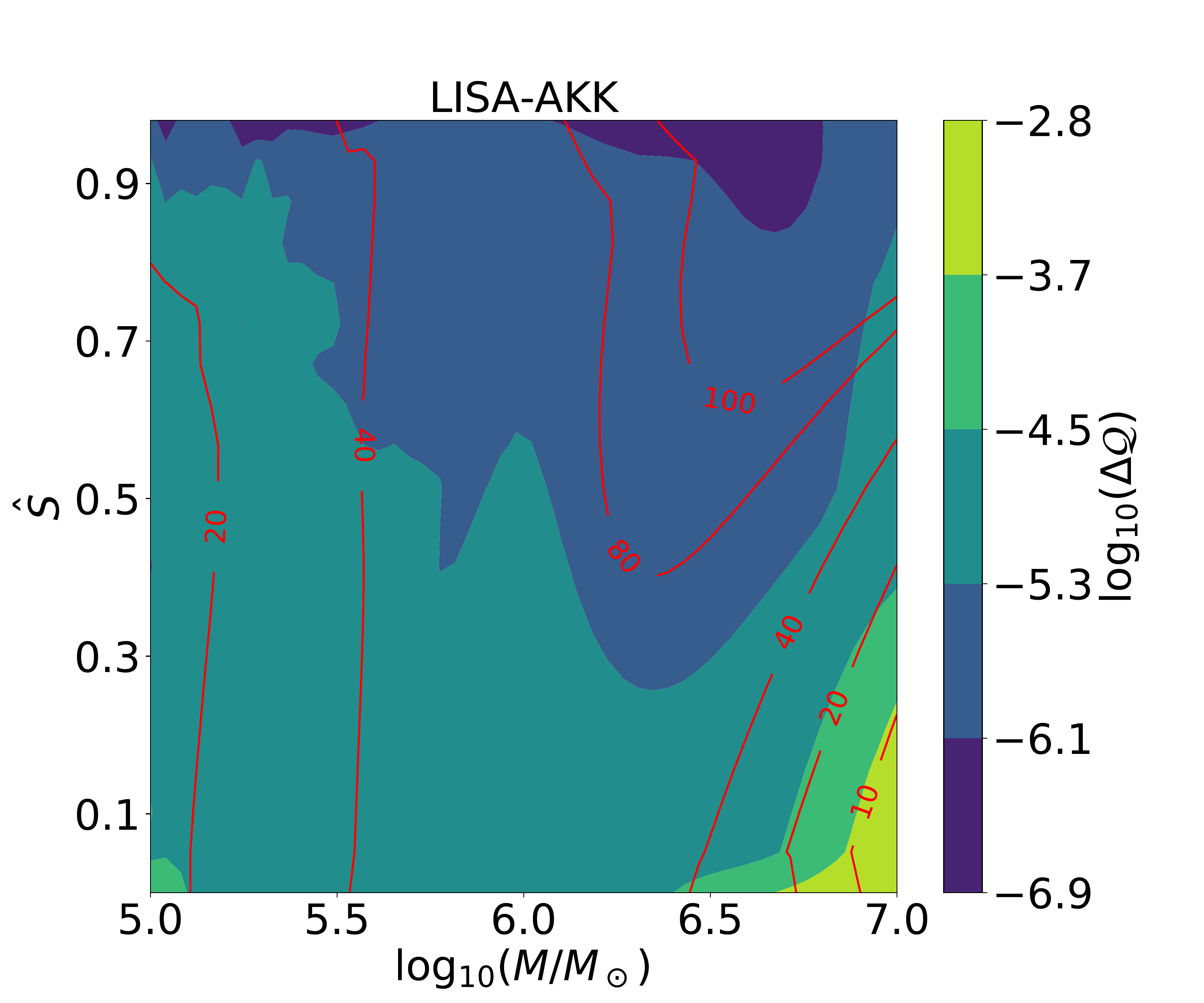}
\caption{Dependence of $\Delta\mathcal{Q}$ on $M$ and $\hat{S}$ using \acp{EMRI} detected with LISA. All parameters are the same as in FIG. \ref{TQcontfig1}.}
\label{LISAcontfig4}
\end{figure*}

\begin{figure*}[!htbp]
	\centering
	\includegraphics[width=0.45\textwidth]{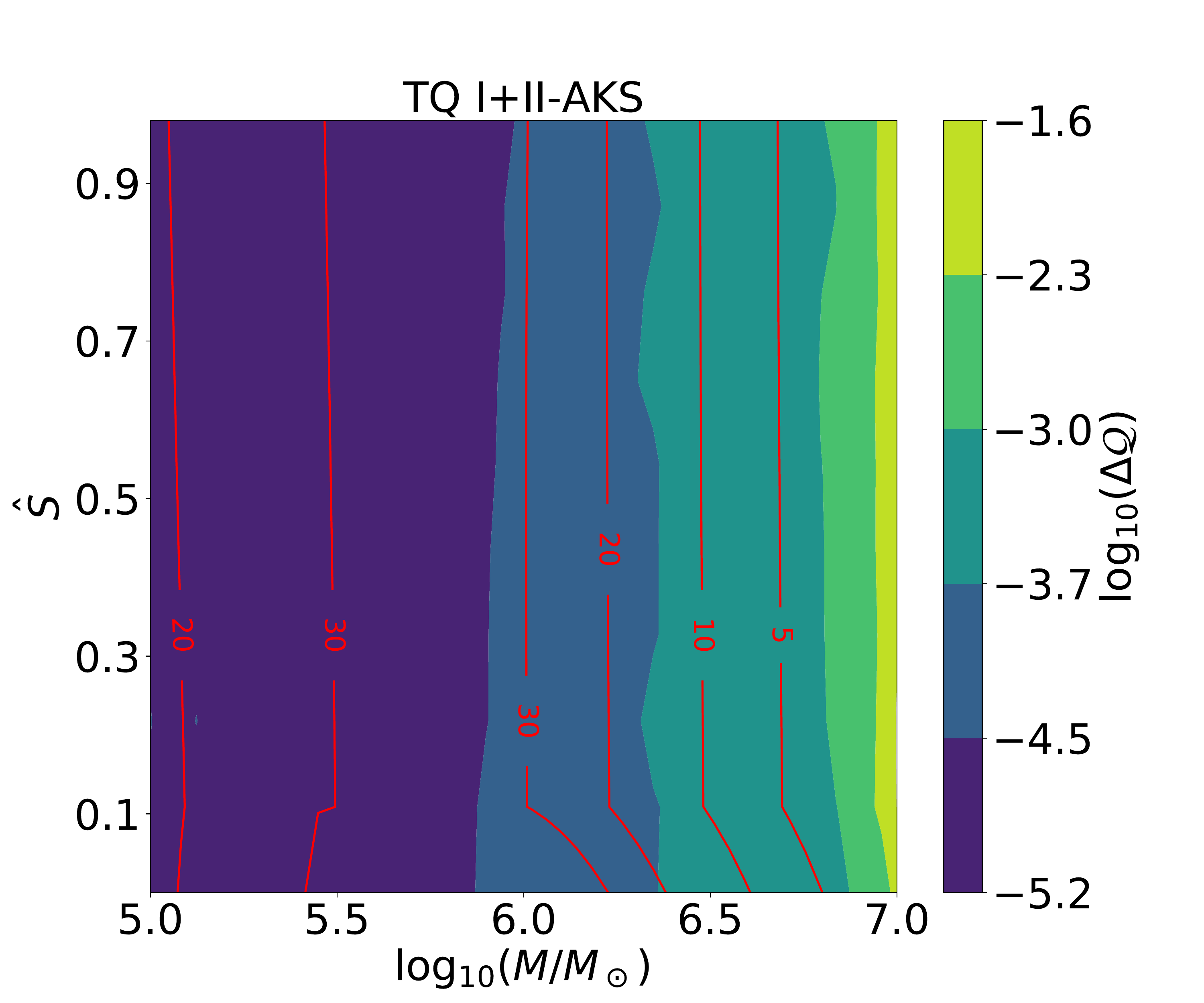}
	\includegraphics[width=0.45\textwidth]{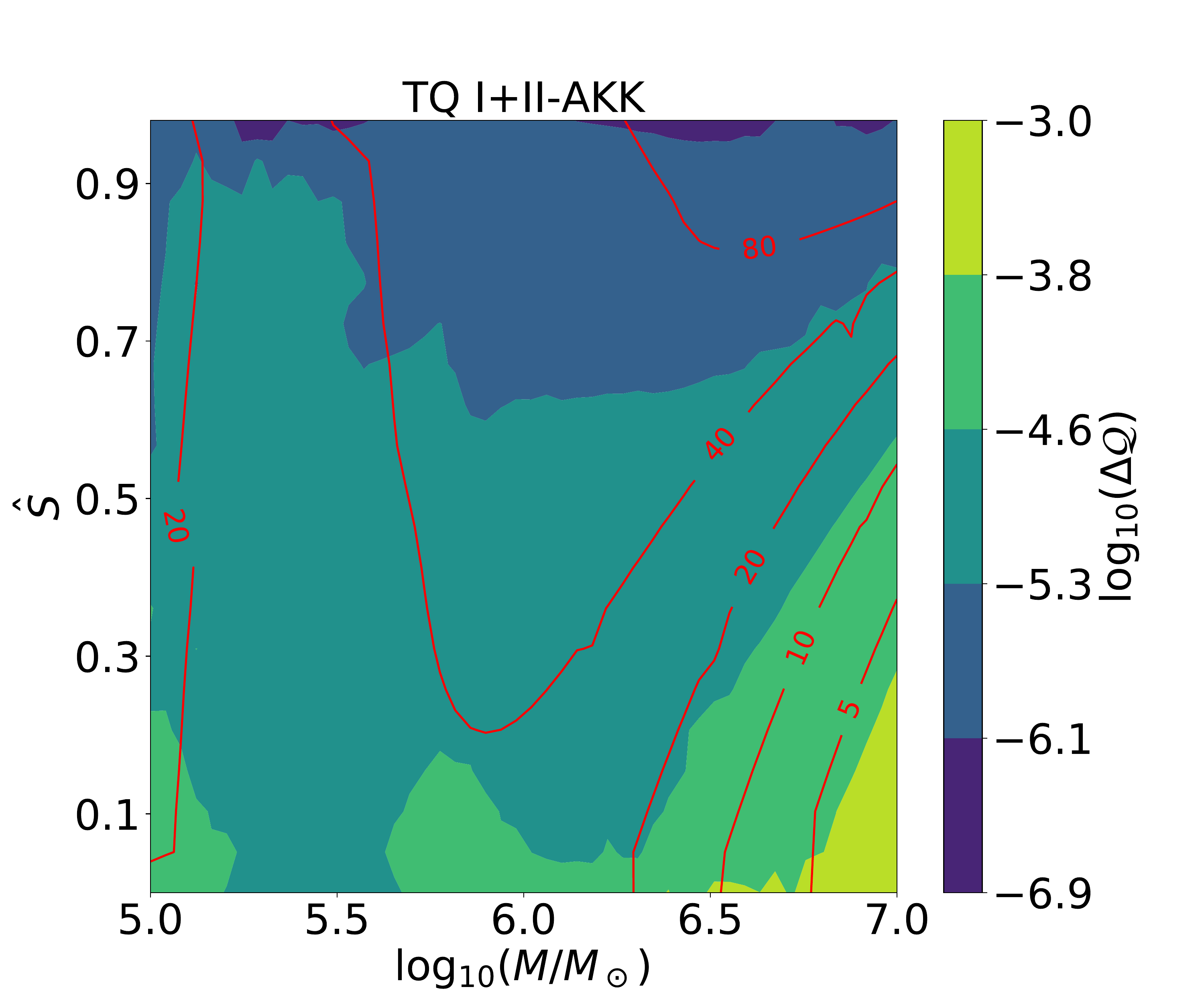}
	\caption{Dependence of $\Delta\mathcal{Q}$ on $M$ and $\hat{S}$ using \acp{EMRI} detected with TQ I+II. All parameters are the same as in FIG. \ref{TQcontfig1}.}
	\label{TQI+IIcontfig2}
\end{figure*}

\begin{figure*}[htbp]
	\centering
	\includegraphics[width=0.45\textwidth]{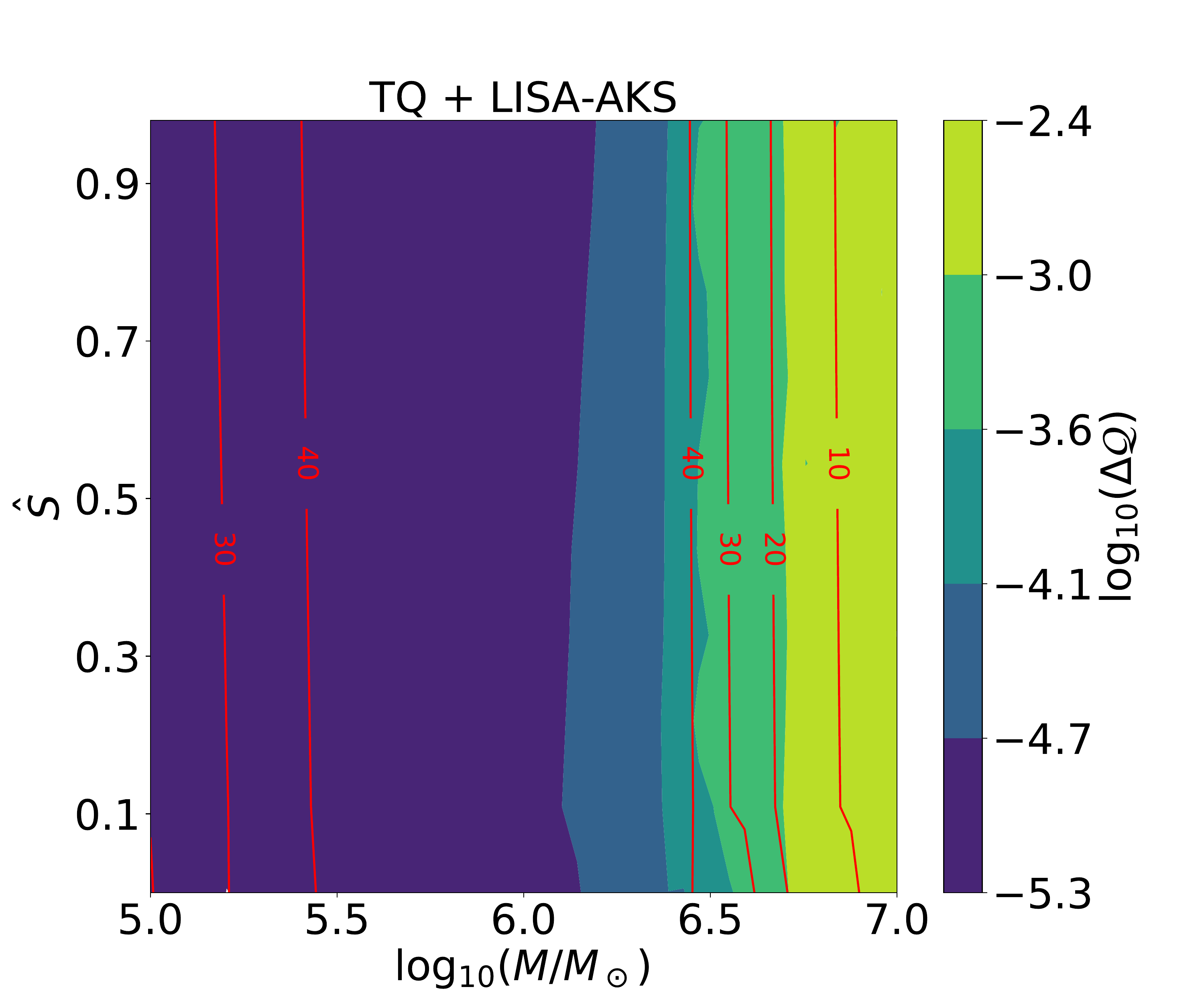}
 \includegraphics[width=0.45\textwidth]{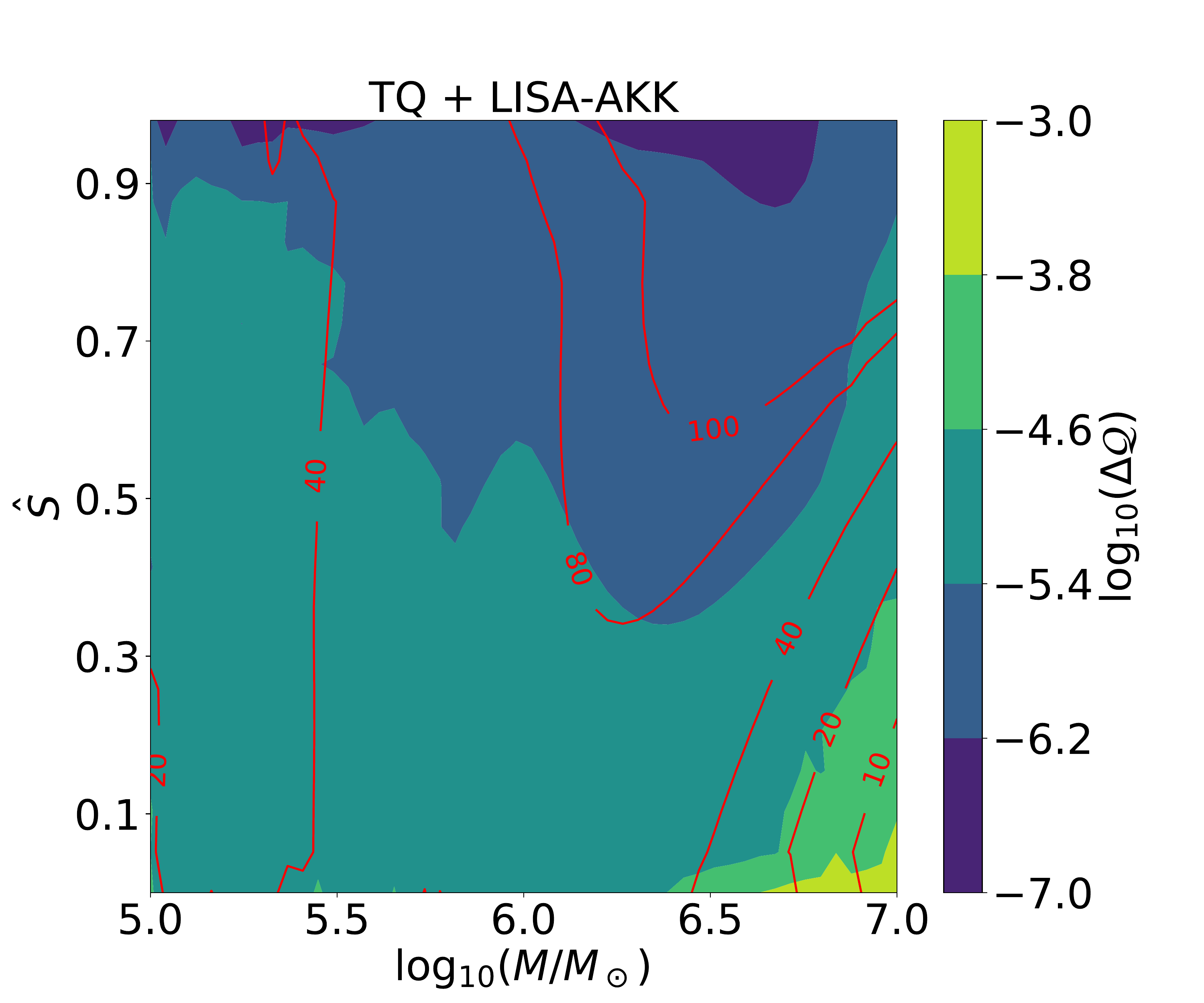}
	\caption{Dependence of $\Delta\mathcal{Q}$ on $M$ and $\hat{S}$ using \acp{EMRI} detected with TQ+LISA. All parameters are the same as in FIG. \ref{TQcontfig1}.}
 \label{TQ+LISAcontfig5}
\end{figure*}

\begin{figure*}[htbp]
	\centering
	\includegraphics[width=0.45\textwidth]{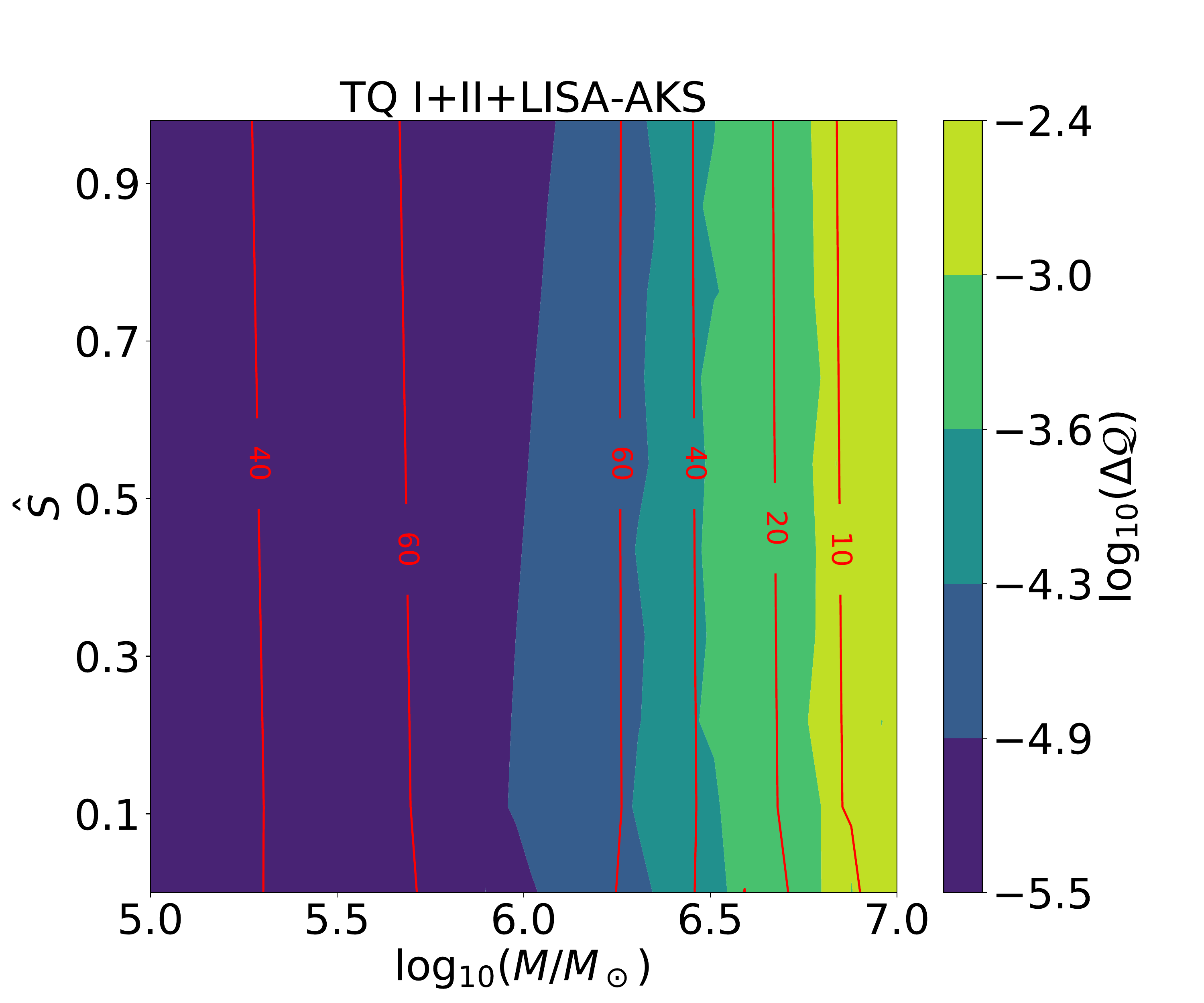}
   \includegraphics[width=0.45\textwidth]{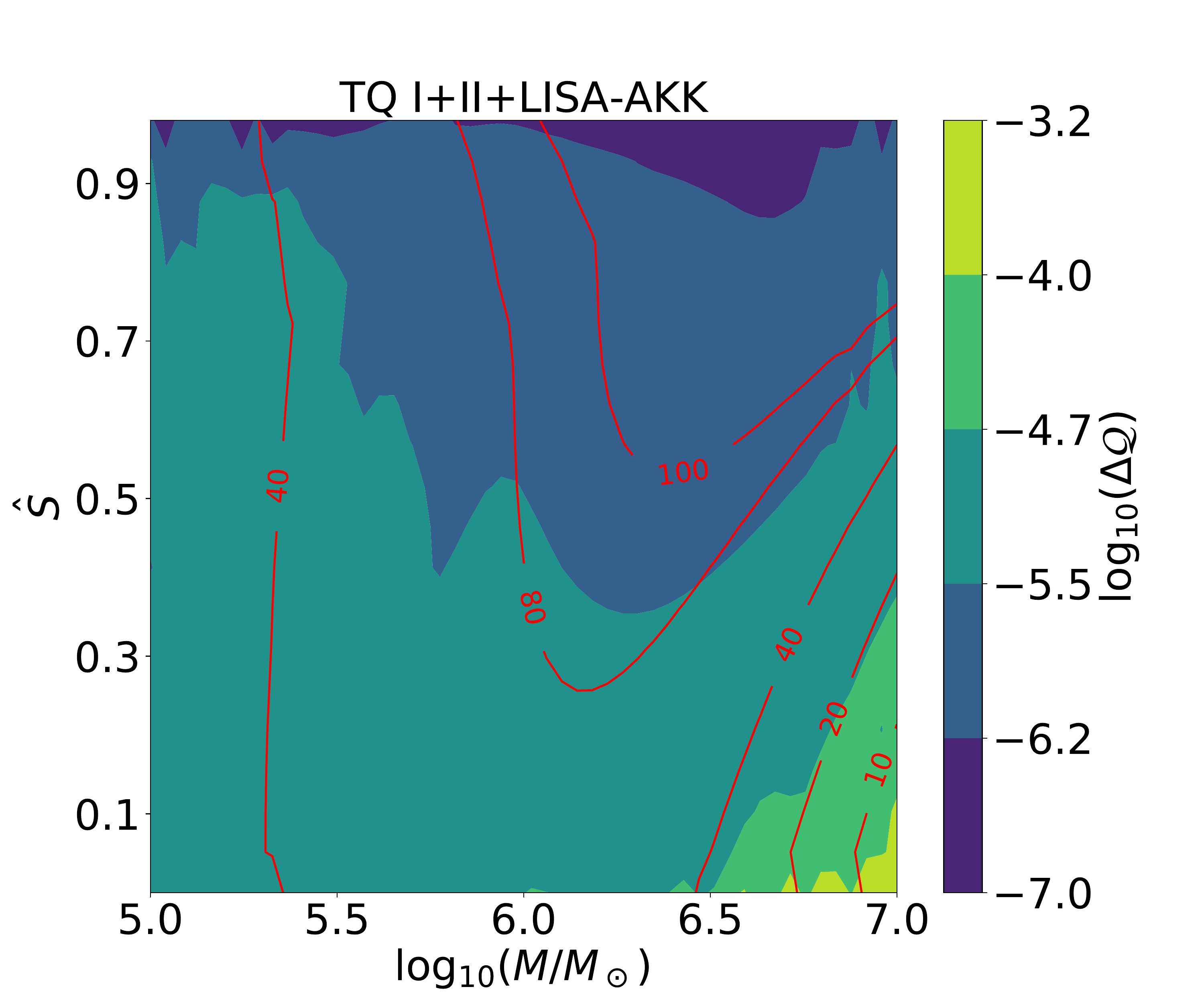}
	\caption{Dependence of $\Delta\mathcal{Q}$ on $M$ and $\hat{S}$ using \acp{EMRI} detected with TQ I+II+LISA. All parameters are the same as in FIG. \ref{TQcontfig1}.}
 \label{TQI+II+LISAcontfig2}
\end{figure*}

\end{document}